\documentclass{article}%
\usepackage{eurosym}
\usepackage[top=3cm, bottom=3cm, left=3cm, right=3cm]{geometry}
\usepackage{amsmath}
\usepackage{amsfonts}
\usepackage{amssymb}
\usepackage{graphicx}%
\begin{document}

\title{Masses of Heavy and Light Scalar Tetraquarks in a Non-Relativistic Quark Model}
\author{Zahra Ghalenovi$^{1}$, Francesco Giacosa$^{2,3}$, and Dirk H.\ Rischke$^{3}$\\$^{1}$Department of Physics, Kosar University of Bojnourd, Bojnourd, Iran\\$^{2}$Institute of Physics, Jan Kochanowski University, 25-406 Kielce, Poland\\$^{3}$Institute for Theoretical Physics, Goethe University, \\Max-von-Laue-Str.\ 1, 60438 Frankfurt, Germany}
\maketitle

\begin{abstract}
\noindent Scalar tetraquark states are studied within the diquark-antidiquark
picture in a non-relativistic approach. We consider two types of confining
potentials, a quadratic and a linear one, to which we also add spin-spin,
isospin-isospin, and spin-isospin interactions. We calculate the masses of the
scalar diquarks and of the ground-state open and hidden charmed and bottom
scalar tetraquarks. Our results indicate that the scalar resonances
$D_{0}^{\ast}(2400)$ and $D_{s}(2632)$ have a sizable tetraquark amount in
their wave function, while, on the other hand, it turns out that the scalar
states $D_{s0}^{\ast}(2317)$ and $X(3915)$ should not be considered as being
predominantly diquark-antidiquark bound states. We also investigate the masses
of light scalar diquarks and tetraquarks, which are comparable to the measured
masses of the light scalar mesons.


\end{abstract}

\section{ Introduction}

One of the most important problems in modern hadron physics is to determine
the structure and the properties of the newly discovered $X,Y,Z$ states as
well other enigmatic mesons, such as $D_{s0}^{\ast}(2317)$, $D_{0}^{\ast
}(2400)$, $D_{s1}^{\ast}(2460)$, etc., see e.g.\ Refs.\ \cite{Godfrey:2008nc,
Swanson:2006st, Brambilla:2010cs,braaten} and refs.\ therein. These states
cannot be accommodated within the simple quark-antiquark picture and are
therefore of special interest.

One possibility is to interpret (some) of these enigmatic mesons as tetraquark
states where the constituent objects are a diquark and an antidiquark. Namely,
although a diquark cannot be a color singlet, the attraction between two
quarks can be strong, as various approaches based on one-gluon exchange
processes \cite{onegluon}, instantons \cite{inst}, lattice calculations
\cite{jaffelatt}, and quark-diquark models for the nucleon \cite{alkofer} and
for baryons in general \cite{Ebert:1996ab} have shown. Thus, the diquark is an
important object for the understanding of baryon structure and is also
potentially important for the understanding of unconventional mesons, most
notably tetraquarks. In particular, in this work we are interested in scalar
diquarks: these are `good diquarks' in Jaffe's terminology \cite{exotica},
with vanishing spin and angular momentum and an antisymmetric flavor wave
function of the type $[q,q^{\prime}]$, where $q,q^{\prime}=u,d,s,c,b$ (a
similar antisymmetric combination is realized in color space).

The masses of heavy tetraquarks as diquark-antidiquark bound states were
studied in the presence of spin-spin interactions in
Ref.\ \cite{Maiani:2005pe,Maiani:2004vq} and later in the comprehensive study
of Ref.\ \cite{Ebert:2010af}. The masses of tetraquarks were also calculated
in a quark model employing a potential derived from the AdS/QCD correspondence
\cite{Carlucci:2007um}, by using a confining interaction and a meson-exchange
potential in a non-relativistic approach \cite{Vijande:2006hj}, by
implementing the Glozman-Riska (flavor-spin) interaction Hamiltonian and
$SU(3)$ flavor symmetry breaking \cite{Jovanovic:2008zza}, and in the
framework of a non-relativistic potential model which includes a three-body
quark interaction \cite{Fan Yong}.

In this paper we continue along these lines and calculate masses of (hidden
and open) charmed and bottom ground-state \emph{scalar} tetraquarks using two
potential models in the non-relativistic limit. As a four-body system, a
tetraquark state is quite different from a conventional $q\bar{q}$ meson and
we solve the problem in a two-step procedure: first, we use a quark-quark
interaction Hamiltonian in order to obtain the mass of a constituent `good
diquark' of the type $[q,q^{\prime}]$. Second, we regard the diquarks as
point-like objects and use a diquark-antidiquark interaction Hamiltonian in
order to obtain the tetraquark masses. In both steps we solve the two-body
Schr\"{o}dinger equation by performing a Taylor expansion \cite{Hassanabadi,
Ozer:2014aja, Koc} or by using a variational method. We compare the values of
the heavy tetraquark masses with the values obtained in previous works and
discuss some possible experimental candidates.

Finally, we focus on the light scalar mesons $f_{0}(500),$ $K_{0}^{\ast
}(800),$ $f_{0}(980),$ and $a_{0}(980)$. These states have been, and still
are, in the center of a vivid debate concerning their nature: there is now a
consensus that they are not predominantly quark-antiquark objects
\cite{pelaez}, but that they emerge either as dynamically generated
molecular-type states, see e.g.\ \cite{molecular,oller,morgan,wolkanowski,gp}
and refs.\ therein, and/or as tetraquark states as proposed some decades ago
by Jaffe \cite{Jaffe:1976ig,Jaffe:1976ih} and further investigated in
Refs.\ \cite{Ebert:1996ab,Maiani:2005pe,
Ebert:2010af,Carlucci:2007um,Vijande:2006hj, Jovanovic:2008zza, Fan Yong,
Jaffe:2004ph,Maiani:2004uc,Gerasyuta:2008ps, Vijande:2003ki, Giacosa:2006rg,
Giacosa:2006tf, Heinz:2008cv, Heupel:2012ua}. (Note that the quark-antiquark
states appear in the spectrum but are heavier, since they lie above 1 GeV
\cite{eLSM1,eLSM2}). We apply the very same two-step approach described above
for a system made of two light diquarks. We evaluate the masses of light
scalar diquarks and tetraquarks and investigate to what extent the light
scalar resonances can be described as scalar tetraquark objects.

The paper is organized as follows. In Sec.\ 2 we introduce the two potential
models and present the methods to solve the Schr\"{o}dinger equation in the
presence of hyperfine interactions. Our predictions for diquarks and scalar
tetraquark masses obtained in the two models are presented and discussed in
Sec.\ 3. Finally, a summary and discussion are presented in Sec.\ 4.

\section{The models}

\subsection{\noindent The Hamiltonian}

The interaction Hamiltonian for the quark-quark interaction leading to the
formation of diquarks is given by%
\begin{equation}
H^{qq}(x)=V^{qq}(x)+H_{hyp}^{qq}\text{ ,} \label{hamiltonian}%
\end{equation}
where the potential $V^{qq}(x)$ consists of three parts:%
\begin{equation}
V^{qq}(x)=V_{conf}(x)-\frac{\tau}{x}-C\text{ .} \label{V(x) interaction}%
\end{equation}
The first term $V_{conf}(x)$ is a confining potential (see the next
subsections), the second term $-\tau/x$ is a Coulomb-like potential due to
one-gluon exchange processes, and $C$ is a constant. The variable $x$ is the
relative quark-quark coordinate. The quantity $H_{hyp}^{qq}$ is the hyperfine
interaction given by:%
\begin{equation}
H_{hyp}(x)=H_{S}(x)+H_{I}(x)+H_{SI}(x)\text{ ,} \label{hamiltonian2}%
\end{equation}
where $H_{S}(x),$ $H_{I}(x)$, and $H_{SI}(x)$ are spin-spin, isospin-isospin,
and spin-isospin interactions, respectively. They read explicitly
\cite{Ghalenovi:2014swa, Ghalenovi:2012zza, Ghalenovi:2012zz,
Ghalenovi:2011zz, Hassanabadi:2008zz, Giannini:2003xx}:%
\begin{equation}
H_{S}=A_{S}\left(  \frac{1}{\sqrt{\pi}\sigma_{s}}\right)  ^{3}\exp\left(
-\frac{x^{2}}{\sigma_{S}^{2}}\right)  (\vec{s}_{1} \cdot\vec{s}_{2})\text{ ,}
\label{HS}%
\end{equation}
\begin{equation}
H_{I}=A_{I}\left(  \frac{1}{\sqrt{\pi}\sigma_{I}}\right)  ^{3}\exp\left(
-\frac{x^{2}}{\sigma_{I}^{2}}\right)  (\vec{t}_{1} \cdot\vec{t}_{2})\text{ ,}
\label{HI}%
\end{equation}%
\begin{equation}
H_{SI}=A_{SI}\left(  \frac{1}{\sqrt{\pi}\sigma_{SI}}\right)  ^{3}\exp\left(
-\frac{x^{2}}{\sigma_{SI}^{2}}\right)  (\vec{s}_{1} \cdot\vec{s}_{2})(\vec
{t}_{1}\cdot\vec{t}_{2})\text{ ,} \label{HSI}%
\end{equation}
where $s_{i}$ and $t_{i}$ are the spin and isospin operators of the $i$-th
quark, respectively, while $A_{k}$ and $\sigma_{k}$ with $k=S,I,SI$ are
constants. Note that the operator $t_{z}$ has eigenvalue $+\frac{1}{2}$ for
the $u$ quark, $-\frac{1}{2}$ for the $d$ quark, and zero for all other quark
flavors. Following Refs.\ \cite{Ghalenovi:2012zza, Ghalenovi:2012zz,
Ghalenovi:2011zz, Hassanabadi:2008zz, Giannini:2003xx}, the spatial dependence
of the hyperfine interaction terms is not modelled by a Dirac $\delta$
function, but by a smooth Gaussian function. The hyperfine Hamiltonian is
treated as a perturbation which slightly modifies the energy levels.

Next, we turn to the diquark-antidiquark potential. First, we recall that the
one-gluon exchange potential is such that the quark-antiquark potential and
quark-quark potentials are related by $V_{q\bar{q}}=2V_{qq}$ (this is due to
the product of Gell-Mann matrices $\vec{\lambda_{i}}\cdot\vec{\lambda_{j}}$,
for details see Refs.\ \cite{SilvestreBrac:1996bg,Blanco:1999qt,
KerenZur:2007vp, Helminen:2000jb,lucha}). When turning to the interaction
between a good diquark and a good antidiquark, we assume the same form as for
a quark-antiquark pair \cite{exotica}. Thus, taking into account the factor
$2$, we get for a diquark-antidiquark system:%
\begin{equation}
H^{D\bar{D}}(x)=V^{D\bar{D}}(x)+H_{hyp}^{D\bar{D}}\text{ ,} \label{hdd}%
\end{equation}
where the potential $V^{D\bar{D}}(x)$ reads%
\begin{equation}
V^{D\bar{D}}(x)=2V_{conf}(x)-\frac{2\tau}{x}-C\text{ .} \label{vdd}%
\end{equation}
The variable $x$ is now the relative diquark-antidiquark coordinate and
$H_{hyp}^{D\bar{D}}$ has the same form as $H_{hyp}$ in
Eq.\ (\ref{hamiltonian2}). When applied to (good) diquarks, the isospin
operator $t_{z}$ has eigenvalue $+\frac{1}{2}$ for the diquark $[u,q]$ (with
$q=s,c,b$), $-\frac{1}{2}$ for the diquark $[d,q]$ (with $q=s,c,b$), and zero
for the diquarks $[u,d]$ and $[q,q^{\prime}]$ (with $q,q^{\prime}=s,c,b$).

\subsection{Quadratic confinement}

In this work, we consider both quadratic and linear potentials in order to
model confining interactions. First, we study the confining potential in
Eq.\ (\ref{hamiltonian}) between two quarks as given by (model 1)%
\begin{equation}
V_{conf}(x)=ax^{2}\text{ ,} \label{Conf potential 1}%
\end{equation}
where $a$ is a positive constant. Since the potential is assumed to depend on
$x$ only, one can factor out the angular part of the two-body wave function.
The remaining radial part of the wave function for the two-body problem with
the unperturbed potential $V^{qq}(x)$ is then determined by the
Schr\"{o}dinger equation%
\begin{equation}
\left[  \frac{d^{2}}{dx^{2}}+\frac{2}{x}\frac{d}{dx}-\frac{l(l+1)}{x^{2}%
}\right]  \psi_{l}(x)=-2m[E_{l}-V^{qq}(x)]\psi_{l}(x)\text{ ,}
\label{equation 1}%
\end{equation}
where $\psi_{l}(x)$ is the radial wave function, $l$ is the angular quantum
number, and $m$ is the reduced mass of the two-body system,%
\begin{equation}
m=\frac{m_{1}m_{2}}{m_{1}+m_{2}}\; , \label{Miu}%
\end{equation}
with $m_{1}$ and $m_{2}$ being the constituent quark (and, subsequently,
diquark) masses. Now we solve the radial Schr\"{o}dinger equation for the
two-body interaction potential (\ref{V(x) interaction}). The transformation%
\begin{equation}
\psi_{l}(x)=x^{-1}\varphi_{l}(x) \label{transformation}%
\end{equation}
reduces Eq.\ (\ref{equation 1}) to the form%
\begin{equation}
\frac{d^{2}}{dx^{2}}\varphi_{l}(x)+\left[  \epsilon_{l}-2max^{2}+\frac{2m\tau
}{x}-\frac{l(l+1)}{x^{2}}\right]  \varphi_{l}(x)=0\text{ .} \label{Phi}%
\end{equation}
The radial wave function $\varphi_{l}(x)$ is a solution of the reduced
Schr\"{o}dinger equation for the wave function of two identical particles with
mass $m$ and interaction potential (\ref{V(x) interaction}), where
\begin{equation}
\epsilon_{l}=2m(E_{l}+C)\text{ .} \label{epsilonl}%
\end{equation}
The effective potential $U_{l}(x)$ reads%
\begin{equation}
U_{l}(x)=2max^{2}-\dfrac{2m\tau}{x}+\dfrac{l(l+1)}{x^{2}}\; .
\label{U function}%
\end{equation}
In order to solve Eq.\ (\ref{Phi}), we perform a Taylor expansion of
$U_{l}(x)$ around $x=x_{l}$,
\begin{equation}
U_{l}(x)\approx U_{l}(x_{l})+\Omega_{l}^{2}(x-x_{l})^{2}\text{ ,}
\label{Taylor exp.}%
\end{equation}
where $x_{l}$ is such that $dU_{l}(x)/dx |_{x=x_{l}}=0$ and
\begin{equation}
\Omega_{l}^{2}=\dfrac{1}{2!}\left.  \dfrac{d^{2}U_{l}(x)}{dx^{2}}\right|
_{x=x_{l}}\text{ .}%
\end{equation}
Substituting Eq.\ (\ref{Taylor exp.}) for $U_{l}$, Eq.\ (\ref{U function}),
into Eq.\ (\ref{Phi}) we find%
\begin{equation}
\frac{d^{2}}{dx^{2}}\varphi_{l}(x)-\Omega_{l}^{2}(x-x_{l})^{2}\varphi
_{l}(x)=-\left[  \varepsilon_{l}-U_{l}(x_{l})\right]  \varphi_{l}(x)\text{ }
\label{new equation}%
\end{equation}
which is the well-known equation for a one-dimensional harmonic oscillator.
Namely, for a particle with mass $m$, oscillation frequency $\omega^{\prime}$,
energy eigenvalues $\varepsilon^{\prime}=\left(  n+\dfrac{1}{2}\right)
\hbar\omega^{\prime}$, and spatial wave function $\phi(x)$, the
one-dimensional harmonic oscillator equation reads:%
\begin{equation}
\dfrac{d^{2}}{dx^{2}}\phi(x)-\frac{m^{2}\omega^{\prime2}x^{2}}{\hbar^{2}}%
\phi(x)=-\dfrac{2m\varepsilon^{\prime}}{\hbar^{2}}\phi(x)\text{ .}
\label{oscillator equation}%
\end{equation}
We consider here the ground state of the scalar diquarks ($l=n=0$). In this
way, upon a comparison of Eq.\ (\ref{new equation}) with
Eq.\ (\ref{oscillator equation}), we have:%
\begin{equation}
\Omega_{0}=\dfrac{m\omega^{\prime}}{\hbar}\text{ , }\ \ \varepsilon_{0}%
-U_{0}(x_{0})=\dfrac{2m\varepsilon^{\prime}}{\hbar^{2}}\text{ .}%
\qquad\label{comparision}%
\end{equation}
Finally, the ground-state energy eigenvalue $E_{0}$ is obtained using
Eq.\ (\ref{epsilonl}):%
\begin{equation}
E_{0,qq}=-C+\dfrac{1}{2m}\left[  U_{0}(x_{0})+\Omega_{0}\right]  \text{ ,}
\label{energy eigenvalues}%
\end{equation}
with the corresponding ground-state wave function
\begin{equation}
\varphi_{0}(x)=\sqrt{\frac{2\Omega_{0}}{\sqrt{\pi}}}e^{-\frac{1}{2}\Omega
_{0}x^{2}}\; , \label{eigenfunction}%
\end{equation}
where the constant term in front is a normalization constant.\newline

The very same mathematical problem needs to be solved for the
diquark-antidiquark state by treating (anti)diquarks as point particles under
the influence of the potential (\ref{vdd}). The energy eigenvalue
$E_{0,D\bar{D}}$ of the tetraquark ground state $n=l=0$ is then calculated in
the same way.

\subsection{Linear confinement}

We also model confinement via a linearly rising potential (model 2):%
\begin{equation}
V_{conf}(x)=ax\text{ .} \label{Conf potential 2}%
\end{equation}
The potential (\ref{V(x) interaction}) is now the well-known Cornell
potential. Similarly to the potential of model 1, we can factorize the angular
part of the Schr\"{o}dinger equation. Upon substituting the potential
(\ref{Conf potential 2}) into Eq.\ (\ref{equation 1}) and using the
transformation (\ref{transformation}) we obtain:%
\begin{equation}
\frac{d^{2}}{dx^{2}}\varphi_{l}(x)+\left[  \epsilon_{l}-2max+\frac{2m\tau}%
{x}-\frac{l(l+1)}{x^{2}}\right]  \varphi_{l}(x)=0\text{ .} \label{Phi 2}%
\end{equation}
We use a variational method to solve the Schr\"{o}dinger equation for the case
$l=0$ using the normalized test function
\begin{equation}
\varphi_{0}(x)=\sqrt{\frac{16p^{3}}{\sqrt{2\pi}}}xe^{-p^{2}x^{2}}\text{ ,}%
\end{equation}
where $p$ is the variational parameter. By minimization of the energy of the
system, we calculate the energy and also the wave function of the system (for
further details of this approach, see Ref.\ \cite{Ghalenovi:2011zz}).\newline
Also in this case, the approach can be easily extended to the calculation of
the ground-state energies of diquark-antidiquark objects.

\section{Diquark and Tetraquark Masses}

In this section we present the results for diquark and tetraquark masses. We
first focus on diquarks and then on the corresponding tetraquarks containing
at least one heavy quark. Finally, we turn to light diquarks and tetraquarks.

\subsection{Diquarks}

The diquark masses obtain a contribution from the constituent quark masses as
well as from the confining and the spin-isospin-dependent interactions:%
\begin{equation}
M_{diquark}=m_{q1}+m_{q2}+E_{0,qq}+\left\langle H_{hyp}\right\rangle \text{ ,}
\label{M-diquark}%
\end{equation}
where $m_{qi}$ is the mass of $i$-th quark and $E_{0,qq}$ is the ground-state
energy calculated in the previous section. The first-order energy correction
from the non-confining potential $\left\langle H_{hyp}\right\rangle $ is
calculated using the unperturbed wave function obtained in Secs.\ 2.2 and
2.3.
\begin{equation}
\left\langle H_{hyp}\right\rangle =\int\mathrm{d}^{3}\mathrm{x}\,
\varphi\,H_{hyp}\,\varphi\text{ .} \label{perturbed energy}%
\end{equation}
For the numerical evaluation, we use for model 1 the light and heavy quark
masses and the parameter $\tau$ from Ref.\ \cite{SilvestreBrac:1996bg}, while
the parameter $a$ and the hyperfine potential parameters are taken from
Refs.\ \cite{Ghalenovi:2012zz, Giannini:2003xx}. In model 2, the light and
heavy quark masses are still taken from Ref.\ \cite{SilvestreBrac:1996bg} but
the potential parameters are from Ref.\ \cite{Valcarce:1995ch}. In both models
the parameter $C$ is obtained by fitting it to the experimental mass of the
$\rho$ meson. The parameters of both models are summarized in Tab.\ 1.

The scalar diquark masses obtained by models 1 and 2 are shown in Tab.\ 2 and
are compared with the theoretical works \cite{Ebert:2010af,
Maiani:2004uc,Chakrabarti}. We note that the predictions of the two models are
similar to each other as well as to previous theoretical calculations.

\begin{table}[ptb]
\caption{Parameters used in our model.}
\begin{center}%
\begin{tabular}
[c]{c|c|c}\hline\hline
Parameter & Model 1 & Model 2\\\hline
$\sigma_{S}$ & 2.87 fm & 2.87 fm\\
$A_{S}$ & 67.4 fm$^{2}$ & 67.4 fm$^{2}$\\
$\sigma_{SI}$ & 2.31 fm & 2.31 fm\\
$A_{SI}$ & --106.2 fm$^{2}$ & --106.2 fm$^{2}$\\
$\sigma_{I}$ & 3.45 fm & 3.45 fm\\
$A_{I}$ & 51.7 fm$^{2}$ & 51.7 fm$^{2}$\\
$m_{u}=m_{d}$ & 277 MeV & 280 MeV\\
$m_{s}$ & 553 MeV & 569 MeV\\
$m_{c}$ & 1816 MeV & 1840 MeV\\
$m_{b}$ & 5206 MeV & 5213 MeV\\
$a$ & 0.73 fm$^{-3}$ & 1.23 fm$^{-2}$\\
$\tau$ & 0.424 & 0.287\\
$C$ & --2.684 fm$^{-1}$ & --2.08 fm$^{-1}$\\\hline\hline
\end{tabular}
\par
\vspace{1cm}
\end{center}
\end{table}

\bigskip

\begin{table}[ptb]
\caption{Diquark masses (in MeV).}
\begin{center}%
\begin{tabular}
[c]{c|ccccc}\hline\hline
Diquark & Model 1 & Model 2 & Ref.\ \cite{Ebert:2010af} &
Ref.\ \cite{Maiani:2004uc} & Ref.\ \cite{Chakrabarti}\\\hline
$\lbrack qq]$ & 406 & 527 & 710 & 395 & 441\\
$\lbrack qs]$ & 678 & 784 & 948 & 590 & 659\\
$\lbrack qc]$ & 1918 & 2012 & 1973 & 1933 & 1980\\
$\lbrack sc]$ & 2147 & 2213 & 2091 & - & 2120\\
$\lbrack qb]$ & 5296 & 5371 & 5359 & - & 5140\\
$\lbrack sb]$ & 5523 & 5563 & 5462 & - & 5210\\\hline\hline
\end{tabular}
\par
\vspace{1cm}
\end{center}
\end{table}

\subsection{Heavy scalar tetraquarks}

Once the diquark masses are calculated, we can evaluate the tetraquark masses
by following the same steps. The explicit expression reads%
\begin{equation}
M_{tetraquark}=m_{diquark}+m_{antidiquark}+E_{0,D\bar{D}}+\left\langle
H_{hyp}^{D\bar{D}}\right\rangle \text{ .} \label{M tetraquark}%
\end{equation}
\newline\begin{table}[ptb]
\caption{Masses of open charmed and bottom tetraquarks (in MeV).}
\begin{center}
\vspace{1cm}
\par%
\begin{tabular}
[c]{cccccc}\hline\hline
Tetraquark & Model 1 & Model 2 & Ref.\ \cite{Ebert:2010af} &
Ref.\ \cite{Maiani:2004vq} & Exp.\ \cite{pdg,Evdokimov:2004iy}\\\hline
$\lbrack cq\bar{q}\bar{q}]$ & 2398 & 2426 & 2399 & - & $D_{0}^{\ast}(2400)$\\
$\lbrack cq\bar{q}\bar{s}]$ & 2618 & 2600 & 2619 & 2371 & $D_{s0}^{\ast
}(2317)$ , $D_{s0}^{\ast}(2632)$\\
$\lbrack cs\bar{q}\bar{s}]$ & 2855 & 2798 & 2753 & - & -\\\hline
$\lbrack bq\bar{q}\bar{q}]$ & 5763 & 5748 & 5758 & - & -\\
$\lbrack bq\bar{q}\bar{s}]$ & 5980 & 5901 & 5997 & - & -\\
$\lbrack bs\bar{s}\bar{q}]$ & 6217 & 6103 & 6108 & - & -\\\hline\hline
\end{tabular}
\end{center}
\end{table}The results for open charmed and bottom tetraquarks are listed in
Tab.\ 3 and compared with other theoretical predictions
\cite{Maiani:2004vq,Ebert:2010af} and experimental candidates
\cite{pdg,Evdokimov:2004iy}. The masses of the tetraquarks are indeed very
similar in all theoretical approaches, with the exception of $[cs\bar{q}%
\bar{s}],$ which in our case turns out to be heavier than in
Ref.\ \cite{Ebert:2010af}. Our results show that the scalar resonance
$D_{0}^{\ast}(2400)$ may contain a sizable tetraquark amount in its wave
function. On the other hand, the resonance $D_{s0}^{\ast}(2317)$ is too light
to be interpreted as a $[cq\bar{q}\bar{s}]$ tetraquark (see also
Ref.\ \cite{walaa} for a discussion concerning conventional quark-antiquark
charmed scalar states). Another interesting but still controversial state is
the so-called $D_{s0}^{\ast}(2632)$ meson observed by SELEX
\cite{Evdokimov:2004iy}, the mass of which fits well to our theoretical predictions.

The results for hidden charmed and bottom tetraquarks are listed in Tab.\ 4
and compared with the theoretical predictions of Refs.\ \cite{Maiani:2004vq,
Ebert:2009zza,Ebert:2007rn,Ebert:2008kb}. Also here, the theoretical results
are compatible with each other.

Quite interestingly, the by now established scalar resonance $X(3915)$ turns
out to be too heavy to be a $cq\bar{c}\bar{q}$ state and too light to be a
$cs\bar{c}\bar{s}$ state. It is then compatible with being a conventional
$\chi_{c0}(2P)$ quarkonium state.

\begin{table}[ptb]
\caption{Masses of double-hidden charmed and bottom scalar tetraquarks and
masses of open charmed and bottom scalar tetraquarks (in MeV).}
\begin{center}%
\begin{tabular}
[c]{ccccc}\hline\hline
Tetraquark & Model 1 & Model 2 & Ref.\ \cite{Ebert:2009zza,Ebert:2008kb} &
Ref.\ \cite{Maiani:2004vq}\\\hline
$\lbrack cq\bar{c}\bar{q}]$ & 3807 & 3662 & 3812 & 3723\\
$\lbrack cq\bar{c}\bar{s}]$ & 4043 & 3862 & 3922 & -\\
$\lbrack cs\bar{c}\bar{s}]$ & 4268 & 4050 & 4051 & -\\\hline
$\lbrack bq\bar{b}\bar{q}]$ & 10521 & 10044 & 10471 & -\\
$\lbrack bq\bar{b}\bar{s}]$ & 10747 & 10228 & 10572 & -\\
$\lbrack bs\bar{b}\bar{s}]$ & 10973 & 10412 & 10662 & -\\\hline\hline
\end{tabular}
\par%
\begin{tabular}
[c]{cccccc}\hline\hline
Tetraquark & Model 1 & Model 2 & Ref.\ \cite{Ebert:2007rn} &  & \\\hline
$\lbrack cq\bar{b}\bar{q}]$ & 7162 & 6908 & 7177 &  & \\
$\lbrack cq\bar{b}\bar{s}]$ & 7399 & 7106 & 7282 &  & \\
$\lbrack cs\bar{b}\bar{q}]$ & 7397 & 7096 & 7294 &  & \\
$\lbrack cs\bar{b}\bar{s}]$ & 7623 & 7285 & 7398 &  & \\\hline\hline
\end{tabular}
\par
\vspace{1cm}
\end{center}
\end{table}

\subsection{Light scalar tetraquarks}

Finally, we apply our formalism to the calculation of the masses of light
tetraquarks. The aim is to understand if the resonances $f_{0}(500),$
$K(800),$ $f_{0}(980),$ and $a_{0}(980)$ contain a sizable tetraquark amount
or not (for experiments concerning these states see
Refs.\ \cite{Aloisio:2002bt, Aitala:2000xu, Ablikim:2005ni,Yao:2006px} and for
theoretical works concerning the tetraquark hypothesis
Refs.\ \cite{Jaffe:1976ig,Jaffe:1976ih,Jaffe:2004ph,Maiani:2004uc,Giacosa:2006rg,
Santopinto:2006my}). In this framework, the scalar diquarks behave under
flavor (and also color) transformations as antiquarks,%
\begin{equation}
\lbrack u,d]\leftrightarrow\bar{s},\qquad\lbrack d,s]\leftrightarrow\bar
{u},\qquad\lbrack s,u]\leftrightarrow\bar{d}\text{ ,} \label{correspondence}%
\end{equation}
therefore one can construct a nonet of tetraquarks where the lightest state is
the $[ud][\bar{u}\bar{d}]$ and corresponds to $f_{0}(500)$, the second
lightest are the kaonic-like states $[sq][\bar{u}\bar{d}],[\bar{s}\bar
{q}][ud]$ (with $q=u,d$ and $I=1/2$) to be identified with $K_{0}^{\ast}(800)$
and, finally, the four tetraquarks ($[sq][\bar{s}\bar{q}]$) with $I=0,1$ which
correspond to $f_{0}(980)$ and $a_{0}(980)$.

Using the parameters of Tab.\ 1 and the diquark masses of Tab.\ 2 generates
tetraquark masses which are 100--200 MeV heavier than the states $f_{0}(500),$
$K_{0}^{\ast}(800),$ $f_{0}(980),$ and $a_{0}(980)$. In order to investigate
whether a better agreement is possible, we re-fit the parameter $C$ of
Eq.\ (\ref{V(x) interaction}) for both models, using the well-known tetraquark
state $a_{0}(980)$ as an input and obtain the light and strange diquark masses
as following: a) model 1 [ $C=-3.507$ fm$^{-1}$ ]: $M_{[qq]}=244$ MeV$,\;
M_{[qs]}=515$ MeV; b) model 2 [ $C=-3.045$ fm$^{-1}$ ]: $M_{[qq]}=330$ MeV$,\;
M_{[qs]}=592$\ MeV. Santopinto and Galata \cite{Santopinto:2006my} have
considered a diquark-antidiquark picture of the light scalar tetraquarks in
the non-relativistic limit, where the masses of the scalar diquarks were
obtained as $M_{[qq]}=275$ MeV$,M_{[sq]}=492$ MeV.

Using the new diquark masses, we obtain the masses of the light scalar
tetraquark nonet listed in Tab.\ 5. Our predictions for the masses of the
light scalar tetraquarks are in good agreement with the experimental data and
also with the results obtained in Refs.\ \cite{Santopinto:2006my} and
\cite{Ebert:2008id}. In our model, a small difference between the masses of
$a_{0}(980)$ and $f_{0}(980)$ arises from the isospin-dependent hyperfine interaction.

\begin{table}[ptb]
\caption{Masses of light tetraquark states (in MeV).}%
\label{double charmed and bottom}
\begin{center}%
\begin{tabular}
[c]{cccccccc}\hline\hline
Resonance & flavor content & $I(J^{p})$ & Model 1 & Model 2 & Ref.
\cite{Santopinto:2006my} & Ref. \cite{Ebert:2008id} & Exp. \cite{pdg}\\\hline
$f_{0}(500)$ & $[ud][\bar{u}\bar{d}]$ & $0(0^{+})$ & 546 & 614 & 550 & 596 &
400--550\\
$K(800)$ & $[ud][\bar{s}\bar{d}]$ & $1/2(0^{+})$ & 765 & 804 & 767 & 730 &
653--701\\
$f_{0}(980)$ & $[us][\bar{u}\bar{s}]+[ds][\bar{d}\bar{s}]$ & $0(0^{+})$ &
962 & 962 & 984 & 992 & 970--990\\
$a_{0}(980)$ & $[us][\bar{u}\bar{s}]-[ds][\bar{d}\bar{s}]$ & $1(0^{-})$ &
984 & 984 & 984 & 992 & 983.5--985.9\\\hline\hline
\end{tabular}
\end{center}
\end{table}

In the context of light scalar states it should be stressed that the role of
loop corrections to the self-energy is surely non-negligible for the masses of
these states \cite{pelaez,oller,morgan,wolkanowski}. Namely, light scalars
have a strong coupling to pseudoscalar mesons and a diquark-antidiquark
configuration can easily transform into a meson-meson one. Moreover, our
approach is non-relativistic, thus its application to the light scalar sector
must be treated with care. Yet, our study shows once more that the light
scalar mesons are not simple quark-antiquark states but may have a sizable
four-quark component. In conclusion, we mention that light scalar mesons also
play an important role at nonzero temperature \cite{Heinz:2008cv} and at
nonzero density \cite{tqfinitedensity}.

\section{Summary}

In this work we have calculated the masses of the ground-state heavy and light
scalar tetraquarks in the framework of a non-relativistic approach with two
types of confining potentials, a quadratically and a linearly rising one, as
well as (iso)spin-(iso)spin interactions. The results for the scalar diquarks
are shown in Tab.\ 2, while the heavy scalar tetraquarks are summarized in
Tabs.\ 3 and 4. The results of both models are compatible with each other,
showing only a mild influence of the particular form of the confining
potential. Moreover, the results are in agreement, apart from a few
exceptions, with previous theoretical calculations of
Refs.\ \cite{Maiani:2005pe,Maiani:2004vq,Ebert:2010af}.

Our results for the masses show that the resonance $D_{s0}^{\ast}(2317)$ is
too light to be predominantly a tetraquark state of the type $cq\bar{q}\bar
{s}$, while the hidden charmed state $X(3915)$ is too heavy to be $cq\bar
{c}\bar{q}$ and too light to be $cs\bar{c}\bar{s}$. On the other hand, the
state $D_{0}^{\ast}(2400)$ and the putative $D_{s0}^{\ast}(2632)$ can contain
an important tetraquark component in their flavor wave function ($cq\bar
{q}\bar{q}$ and $cq\bar{q}\bar{s},$ respectively). In addition to already
existing experimental candidates, we also made predictions for the masses of
scalar tetraquark states which can be discovered in the future (see Tabs.\ 3
and 4). Namely, some of the $X,Y,$ and $Z$ states could turn out to be scalar
objects. Finally, we have also studied the light scalar sector of QCD and
found that the masses of light scalar mesons $f_{0}(500),$ $K_{0}^{\ast
}(800),$ $f_{0}(980),$ and $a_{0}(980)$ can be described well in the
tetraquark picture (see Tab.\ 5).

In this work the masses of the tetraquarks are calculated by using a static
approach. Mass shifts take place as soon as interactions and quantum
fluctuations are taken into account. These modifications are typically small
for hadrons which are (i) narrow and (ii) are far from any decay threshold.
For what concerns point (i), the ratio $\Gamma/(M-E_{th})$, where $\Gamma$ is
the decay width, $M$ the mass of an hadron, and $E_{th}$ the lowest decay
threshold of the state, is an important quantity to estimate the role of loops
\cite{gp}. This ratio is indeed large for the light scalar mesons $f_{0}(500)$
and $K_{0}^{\ast}(800)$ (see the discussion Sec. 3.3), thus loop corrections
are surely also an important element towards their understanding. Even when
this ratio is relatively small (as it usually is for mesons in the charmonium
region), one should then consider point (ii): namely, when $M$ is close to one
of the decay threshold (not necessarily the lowest), distortion of the
spectral functions, mass shifts and sizable meson-meson amounts in the
wave-function of the unstable meson may occur. This is the case of the light
scalar mesons $f_{0}(980)$ and $a_{0}(980)$: both of them are fairly distant
from the lowest threshold ($\pi\pi$ and $\pi\eta$ respectively), but very
close to the $KK$ threshold. Similarly, the state $D_{s0}^{\ast}(2317)$ is
pretty close to the $DK$ threshold. More in general, many of the newly
discovered $X,Y,$ and $Z$ resonances are close to one of their intermediate
threshold, thus care is definitely needed since the role of loops can be very important.

In view of this discussion, it must be stressed that also the calculation of
decay widths should be performed in the future. Namely, it is possible that
some of the predicted tetraquark states are, due to a `fall apart' decay
mechanics, \textit{too wide} to be measured and that therefore will never be
seen in experiments. This possibility would explain why only some of the (many
possible) tetraquark states are actually detectable, that is when the energy
threshold of the main decay channel is not too far from the mass of the
tetraquark state, in such a way that the kinematic suppression balances the
large decay amplitude. Indeed, this pattern takes place for the light scalar
mesons, where $f_{0}(500)$ and $K_{0}^{\ast}(800)$ are very broad, while
$f_{0}(980)$ and $a_{0}(980)$ are narrow due to the nearby kaon-antikaon threshold.

Another (indeed related) improvement is to go beyond the two-step calculations
performed in this work. Surely, it is much easier to solve two two-body
problems than a four-body problem, but a general feature of our model (as well
as of other tetraquark models) is that the diquarks have a dimension which is
comparable to that of the tetraquark (about $1$ fm). In this respect, there
are also strong quark-antiquark correlations within the tetraquark, because
the diquarks cannot be considered as point-like objects. Just as mentioned
above, the interchange of a diquark-antidiquark $(qq)(\bar{q}\bar{q})$ bound
state with a more molecular-like quark-antiquark $(q\bar{q})(q\bar{q})$ surely
takes place (and is related to the decay of the tetraquark in ordinary
mesons). Thus, the view of a pure diquark-antidiquark bound state serves as a
simple (albeit useful) approximation of the problem, but in the future one
should also go beyond it and solve a (relativistic) four-body problem.

In addition to the listed needed improvements, we also mention that our
approach can be extended to other quantum numbers as well, thus being
potentially interesting to investigate further up to now not yet understood mesons.

\end{document}